# Dynamic Opening of a Gap in Dirac Surface States of the Thin-Film 3D Topological Insulator $Bi_2Se_3$ Driven by the Dynamic Rashba Effect

Yuri D. Glinka [a,b, *], Tingchao He [c, *], Xiao Wei Sun [a,d, *]

[a] Guangdong University Key Lab for Advanced Quantum Dot Displays and Lighting, Shenzhen Key Laboratory for Advanced Quantum Dot Displays and Lighting, Department of Electrical and Electronic Engineering, Southern University of Science and Technology, Shenzhen 518055, China
[b] Institute of Physics, National Academy of Sciences of Ukraine, Kiev 03028, Ukraine
[c] College of Physics and Energy, Shenzhen University, Shenzhen 518060, China
[d] Shenzhen Planck Innovation Technologies Pte Ltd., Longgang, Shenzhen 518112, China

Optical control of Dirac surface states (SS) in topological insulators (TI) remains one of the most challenging problems governing their potential applications in novel electronic and spintronic devices. Here, using visible-range transient absorption spectroscopy exploiting ~340 nm (~3.65 eV) pumping, we provide evidence for dynamic opening of a gap in Dirac SS of the thin-film 3D TI $Bi_2Se_3$, which has been induced by the dynamic Rashba effect occurring in the film bulk with increasing optical pumping power (photoexcited carrier density). The observed effect appears through the transient absorption band associated with inverse-bremsstrahlung-type free carrier absorption in the gapped Dirac SS. We have also recognized experimental signatures of the existence of the higher energy Dirac SS in the 3D TI $Bi_2Se_3$ (in addition to those known as SS1 and SS2) with energies of ~2.7 and ~3.9 eV (SS3 and SS4). It is evidenced that the dynamic gap opening has the same effect on the Dirac SS occurring at any energy.

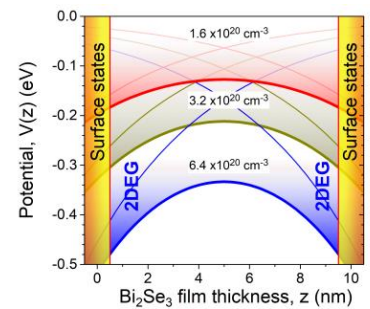

Dirac surface states (SS) in the 3D topological insulator (TI) $Bi_2Se_3$ emerge because of strong spin-orbit interaction while preserving time-reversal symmetry.[1,2] Owing to linear dispersion, they are characterized by two Dirac cones joined through the helical Dirac point (a node), thus forming gapless Dirac SS. The corresponding low-energy quasiparticles nearby a node are known as massless helical Dirac fermions that govern metal-like surface conductivity in the 3D TI $Bi_2Se_3$, despite the insulating nature of gapped bulk states (BS) (bandgap $E_g$ ~0.3 eV). Another interesting feature of Dirac SS is spin-momentum locking, providing electrons with the opposite spin to move into opposite directions.[3] However, all these unique properties are expected to disappear with a topological quantum phase transition introducing an energy gap between the Dirac cones and transferring the quantum system into a trivial phase.[4-6] The effect of gapping the Dirac SS of the 3D TI $Bi_2Se_3$ is hence a crucial aspect of studying these materials with regard to their potential applications in novel electronic and spintronic devices. In addition to the low-energy Dirac SS in the midgap of $Bi_2Se_3$ (SS1), which are usually partially occupied due to natural *n*-doping, there exist unoccupied higher energy Dirac SS. However, only the existence of higher energy Dirac SS with energy ~1.5 eV (SS2) has been evidenced experimentally and theoretically so far.[7]

Because the magnetic field exclusively breaks time reversal symmetry, nonmagnetic perturbations cannot open a gap at the Dirac SS nodes.[2] Consequently, this effect can be achieved in ultrathin $Bi_2Se_3$ films when the opposite-surface spin-polarized Dirac SS interact to each other through magnetic coupling.[4] The critical thickness for this intersurface magnetic coupling has been found to be ~6 quintuple layers (QL) (QL ~ 0.954 nm). This limit introduces a natural static divide between the 3D and 2D TI $Bi_2Se_3$, thus being of different topological phases. Alternately, one can use ultrashort mid-IR pulses of circularly polarized light as breaking time-reversal symmetry, like a magnetic field, to study the dynamic gap opening in Dirac SS of the 3D TI $Bi_2Se_3$.[8] Furthermore, this kind of direct optical control seems to be not necessarily unique, since similar pulsed light-induced topological phase transitions have also been demonstrated to occur in Dirac and Weyl semimetals through the indirect effects, such as coherent phonon generation and magnetization screening by photoexcited carriers.[9,10] The latter dynamic indirect effects have been found to persist as long as the photoexcited electron population remained sufficiently high.

In this letter we report on another example of dynamic indirect opening a gap at the Dirac SS node in the ~10 QL thick 3D TI

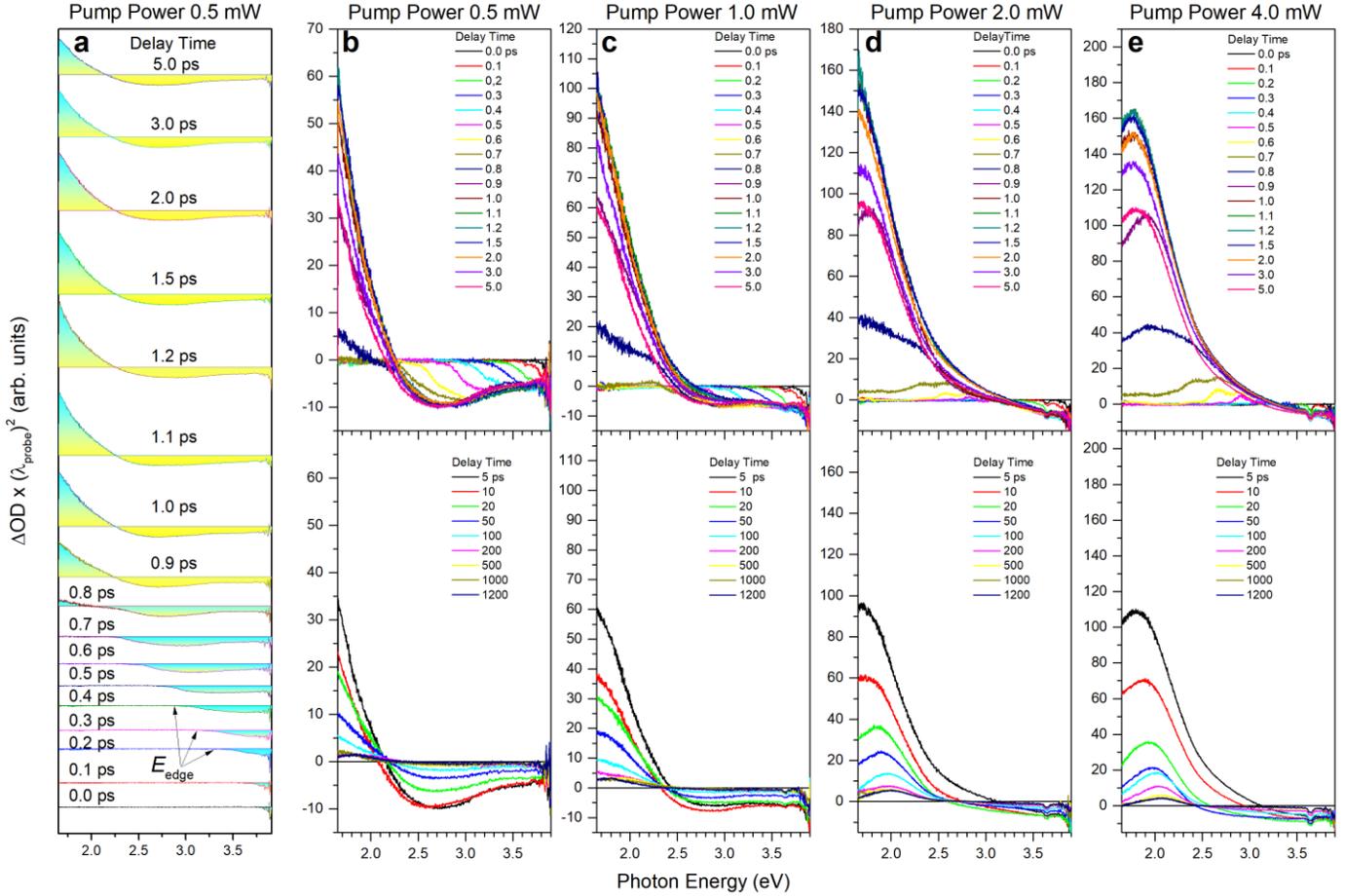

**Figure 1.** TA spectra of the 2D TI $Bi_2Se_3$. (a) - (e) Set of TA spectra of the 4 QL thick $Bi_2Se_3$ film measured at delay times indicated by the corresponding colors using the ~340 nm pumping (~3.65 eV photon energy) of different pumping power, as indicated for each of the columns. (a) and (b) shows the same TA spectra for clarity. The zero-intensity lines of TA spectra in (a) were shifted along the ΔOD axis (optical density change) for better observation. The factor $(\lambda_{probe})^2$ in the ΔOD axis arises due to wavelength-to-energy units' transformation. The low-energy edge of the transiently excited electron population ($E_{edge}$) is indicated in (a).

$Bi_2Se_3$ film driven by the dynamic Rashba effect occurring in the film bulk with increasing optical pumping power (photoexcited carrier density, $n_e$). The ~4 QL thick 2D TI $Bi_2Se_3$ film has been used as a reference sample. Both films were grown on the 0.3 mm $Al_2O_3$(0001) substrates using molecular beam epitaxy, being capped afterward with a 10 nm thick $MgF_2$ protecting layer (Supporting Information). The effect has been observed using visible-range transient absorption (TA) spectroscopy (Supporting Information), the technique that allows directly probing the transiently excited electron population through the Pauli blocking and inverse-bremsstrahlung-type free carrier absorption (FCA) mechanisms.[11] We have exploited one-photon linearly polarized pumping with photon energy ~3.65 eV. As a result, we were able to identify experimental signatures of the existence of the higher energy Dirac SS with energies of ~2.7 and ~3.9 eV (SS3 and SS4). We have also shown that despite the non-magnetic nature of photoexcited carrier, their accumulation in the surface quantum wells (SQW) creates 2D electron gas (2DEG)[12,13,14] in a close vicinity of the film surfaces, which due to Rashba spin-splitting[13,15] introduces the corresponding magnetic moment. Consequently, the effect has been observed with increasing photoexcited carrier density when the surface potential gradient becomes high enough to generate a significant out-of-plane spin polarization. The resulting net magnetization of the 2DEG results in the dynamic gap opening in Dirac SS.[5,6]

Figure 1 shows TA spectra of the 2D TI $Bi_2Se_3$ (4 QL thick film) measured at certain delay times using different pumping powers. In general, the spectra are quite similar to those measured with two-photon pumping exploiting ~1.7 eV photons and demonstrate specific spectral shapes being completely governed by the product of the corresponding density of states and the Fermi-Dirac distribution.[11] Specifically, for the lowest pumping power applied [~0.5 mW ($n_e$ ~1.5 × $10^{20}$ $cm^{-3}$)], the initial stage of the relaxation dynamics (0.0 to 0.7 ps) is dominated by the negative contribution associated with the conduction band (CB) absorption bleaching (AB) occurring in the higher energy bulk states and Dirac SS (CB-AB).[11,16] Additionally, starting from ~0.8 ps, the positive contribution begins to develop, being maximized within ~1.2 ps and associated with the inverse-bremsstrahlung-type FCA in the gapped Dirac SS2.[11] Finally, both contributions decay within the much longer timescale of ~200 ps (Supporting Information, Figure S1).

Upon increasing the pumping power, the positive contribution (FCA) progressively increases and extends toward the higher energy range, occupying the entire visible region, remarkably reducing the negative contribution (CB-AB), and finally revealing a maximum at ~2.0 eV. This behavior clearly points out that the relaxation of the transiently excited electron population with increasing the photoexcited carrier density is predominantly appeared through the gapped Dirac SS. Specifically, the resulting transient accumulation of electrons in the upper Dirac cone of the gapped Dirac SS2 increases the local carrier density and hence appears through the inverse-bremsstrahlung-type FCA that peaked near ~2.0 eV.[11,16] For the maximal pumping power applied in our measurements [~4.0 mW ($n_e$ ~1.2 × $10^{21}$ cm$^{-3}$)], the CB-AB contribution becomes negligible, and the higher energy edge of the positive contribution directly images the higher energy tail of the Fermi-Dirac distribution in the gapped Dirac SS2.[11]

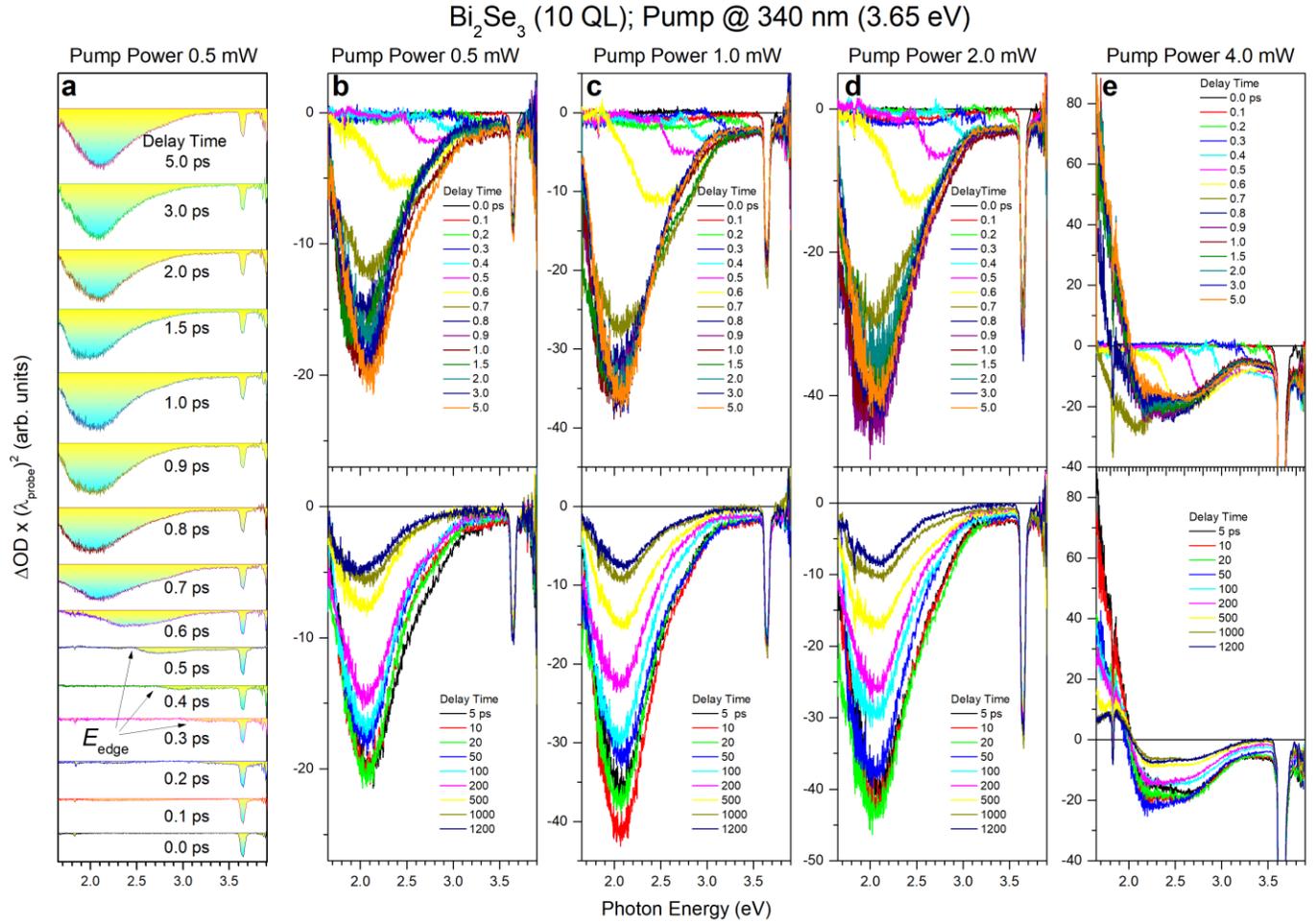

**Figure 2.** TA spectra of the 3D TI $Bi_2Se_3$. (a) - (e) Set of TA spectra of the 10 QL thick $Bi_2Se_3$ film measured at delay times indicated by the corresponding colors using the ~340 nm pumping (~3.65 eV photon energy) of different pumping power, as indicated for each of the columns. (a) and (b) shows the same TA spectra for clarity. The zero-intensity lines of TA spectra in (a) were shifted along the ΔOD axis (optical density change) for better observation. The factor $(\lambda_{probe})^2$ in the ΔOD axis arises due to wavelength-to-energy units' transformation. The low-energy edge of the transiently excited electron population ($E_{edge}$) is indicated in (a). The narrow features peaked at ~3.65 and ~1.825 eV are the VB-AB response (see the text) and its first order diffraction grating replica of the spectrometer that is initially designed for spectroscopic measurements using the second order diffraction maximum.

On the contrary, TA spectra of the 3D TI $Bi_2Se_3$ (10 QL thick film) reveal exclusively the negative contribution (CB-AB) for low and moderate pumping powers, whereas the positive contribution (FCA) manifest itself only at maximal pumping power applied (~4.0 mW) (Figure 2). The overall electron relaxation dynamics of the negative and positive contributions for the 3D TI $Bi_2Se_3$ occurs within the longer timescale exceeding ~1.0 ns (Supporting Information, Figure S2), as observed also with two-photon pumping exploiting ~1.7 eV photons.[11] It is worth noting that there exists another narrow negative contribution that peaked at ~3.65 eV, exactly matches with the pumping photon energy, and is constantly presented in all TA spectra, regardless of delay time. Consequently, this constant feature can be associated with the AB response induced by the one-photon pumping of the valence band (VB) electrons to the CB (the process that is known as VB-AB) [Figure 3(a) and (b)].[11,16] The latter feature characterizes mainly the slow relaxation of holes in the $Bi_2Se_3$ bulk and is less informative.[11]

Because the negative contribution (CB-AB) of TA spectra images the relaxation of the transiently excited electron population in the higher energy bulk states and gapless Dirac SS through the longitudinal optical (LO)-phonon cascade emission, including LO-phonon-assisted or disorder-assisted scattering via the Dirac point, whereas the positive contribution (FCA) is an indication of the electron accumulation in the upper Dirac cone of the gapped Dirac SS2,[11] we associate the pumping power effect observed in the 3D TI $Bi_2Se_3$ with dynamic opening a gap at the Dirac SS2 node. We note here that such an effect has not been observed when the two-photon pumping regime exploiting ~1.7 eV photons was applied,[11] probably due to the lower efficiency of two-photon pumping. It is also worth noting that the observed effect has the dynamic nature and is not accompanied by any damage to the sample. The positive contribution in TA spectra of the 3D TI $Bi_2Se_3$ can hence be switched on/off multiple times by changing the optical pumping power.

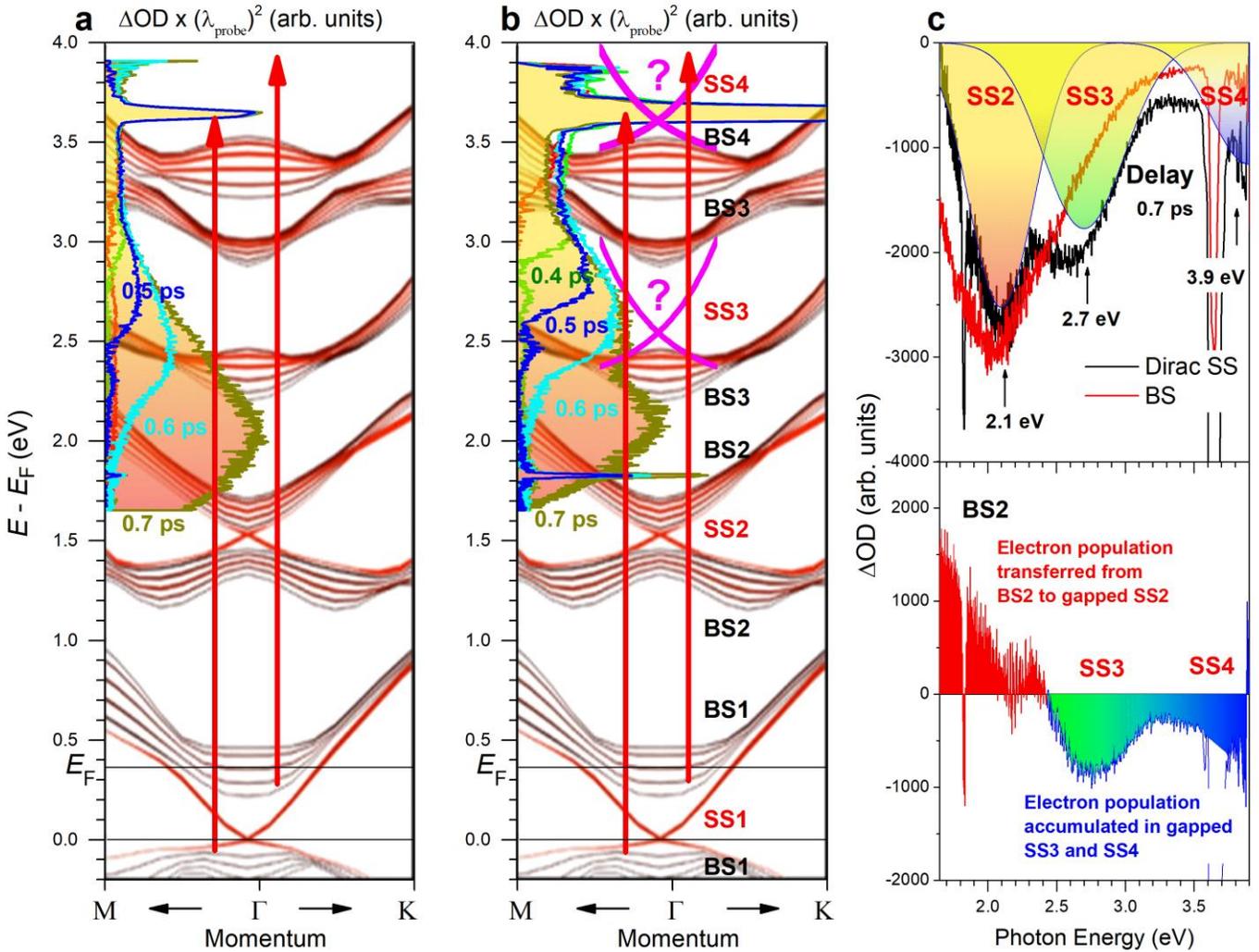

**Figure 3.** TA spectra of the 3D TI $Bi_2Se_3$ measured below and above the critical power. (a) and (b) Set of TA spectra of the 10 QL thick $Bi_2Se_3$ film measured within the 0.3 to 0.8 ps delay timescale using the ~340 nm pumping (~3.65 eV photon energy) of two pumping powers being below (~2.0 mW) and above (~4.0 mW) the critical power, respectively [the same TA spectra as those shown in Fig. 2(d) and (e)]. The spectra are superimposed on the electronic band structure of the 6 QL thick $Bi_2Se_3$ film calculated in Ref. 7. The pumping transitions (red vertical up arrows) originating from the VB states and from the upper Dirac cone of SS1 below the Fermi energy ($E_F$) are shown. The bulk states and the Dirac surface states are marked as BS and SS, respectively. The predicted higher energy Dirac SS are shown in (b). (c) Two spectra measured at 0.7 ps delay time from (a) and (b) corresponding to the dominant BS relaxation and to the dominant Dirac SS relaxation (upper panel) and their difference (lower panel). The fit to the data for the dominant Dirac SS relaxation using Gaussian profiles is shown in the upper panel, whereas the extracted contributions corresponding to the electron population transferred from BS2 to SS2 and to that accumulated in the gapped Dirac SS3 and SS4 is shown in the lower panel.

Figure 3 (a) and (b) show a part of TA spectra of the 3D TI $Bi_2Se_3$ shown in Figure 2, which were measured using two pumping powers corresponding to those being below and above the critical power for the dynamic topological phase transition. The spectra were superimposed on the electronic band structure of the 6 QL thick $Bi_2Se_3$ film,[7] which still applies to the 3D TI $Bi_2Se_3$. One can see that for all pumping powers being below the critical value, the transiently excited electron population gradually and monotonically extends toward the lower energy side, thus demonstrating a spectacular cooling of photoexcited electrons temporally occupying all available states through the LO-phonon cascade emission.[11] The monotonic character of the relaxation dynamics points out that the corresponding rate is quite constant although different higher energy bulk states and Dirac SS are involved. The latter behavior also suggests that LO-phonon-assisted scattering via the Dirac point dominates over the disorder-assisted relaxation mechanism. The occupational dynamics progressively covers the entire visible range, being finally maximized within 0.7 ps at ~2.0 eV. The latter spectral position is known to image the accumulation of electrons in the upper Dirac cone of the gapless Dirac SS2, whereas the transiently excited population contributing at ~1.65 eV images the accumulation of electrons in the bulk states (BS2).[11]

Once the pumping power exceeds the critical value, the relaxation dynamics changes dramatically as a consequence of the dynamic topological phase transition. First, the BS2 population drops down significantly within ~0.7 ps [Figure 3(b)] due to the LO-phonon-assisted vertical transport of electrons towards the upper Dirac cone of the gapped Dirac SS2.[11] The corresponding transiently accumulated electron population peaked at ~2.1 eV, thus being slightly higher than that in the gapless Dirac SS2 (~2.0 eV). Second, two additional contributions in TA spectra peaked at ~2.7 and ~3.9 eV gradually develop within ~0.7 ps. These contributions unambiguously point to the termination of the monotonic relaxation, resulting also in the transient accumulation of electrons in the higher energy states. Because this rearrangement occurs simultaneously with opening of a gap at the Dirac SS2 node and because the additional contributions peaked between the corresponding higher energy bulk states, we associate this observation with the transient accumulation of electrons in the upper Dirac cone of the higher energy gapped Dirac SS (SS3 and SS4) [Figure 3(b)], as has recently been predicted.[11] This relaxation dynamics also suggests that the higher energy Dirac SS (SS3 and SS4) exist permanently at any pumping power applied, however, they manifest themselves exclusively when the gap in Dirac SS is opened. The additional contributions can be extracted explicitly by taking the difference between the two TA spectra measured using pumping powers being below and above the critical pumping power [Figure 3(c)]. This observation suggests that the dynamic gap opening effect equally influences all Dirac SS occurred at any energies. It is worth noting that the actual peak in TA spectra associated with SS4 is most likely located at energies slightly above 3.9 eV, which are not achievable in our experiments.

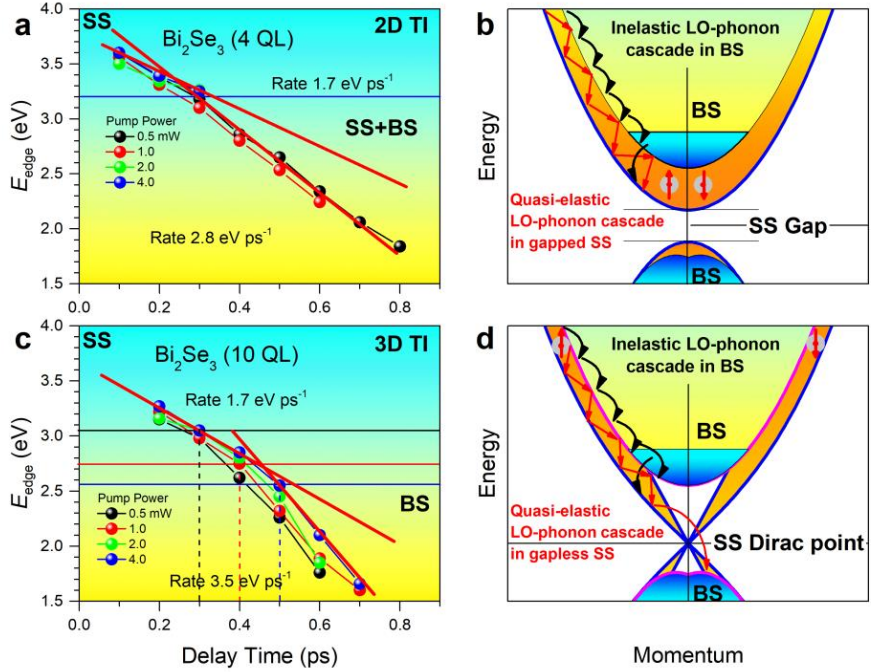

**Figure 4.** Carrier relaxation rates in the 2D and 3D TI $Bi_2Se_3$. (a) and (c) The temporal evolution of the low-energy edge of transiently excited electron population [$E_{edge}$, as defined in Fig. 1(a) and 2(a)] for the 2D and 3D TI $Bi_2Se_3$, respectively, measured using different pumping powers, as indicated by the corresponding colors. The linear fits (the red color straight lines) and the corresponding electron energy loss rates in eV ps$^{-1}$ units are shown for the higher energy Dirac SS and the bulk states (BS), as indicated. The horizontal solid lines and vertical dashed lines in (c) indicate the energy and delay time, at which the electron energy loss rate changes. (b) and (d) The schematic representation of the gapped and gapless Dirac SS, respectively. The different types of LO-phonon cascade emission in BS (inelastic) and Dirac SS (quasi-elastic) and the BS-to-SS vertical electron transport are shown. The LO-phonon-assisted electron scattering via the SS Dirac point is also shown in (d).

The corresponding rates of electron energy loss through the LO-phonon cascade emission in the 2D and 3D TI $Bi_2Se_3$ can be estimated by analyzing the temporal shift of the low-energy edge of the transiently excited electron population [$E_{edge}$, as defined in Figures 1(a) and 2(a)] within ~0.8 ps, thus within the characteristic time after which the relaxing electrons start accumulating in the upper Dirac cone of the Dirac SS2 [Figure 3(a) and (b)]. The resulting electron relaxation dynamics [Figure 4(a) and (c)] is quite similar to that observed with two-photon pumping exploiting ~1.7 eV photons.[11] It reveals that the two-stage electron cooling behavior is associated with the relaxation of photoexcited electrons through the LO-phonon cascade emission initially occurring in the higher energy Dirac SS (presumably SS4) and switching subsequently to the higher energy bulk states (presumably BS4, BS3, and BS2) and finally to the Dirac SS again (SS2). The overall relaxation dynamics includes hence the SS-bulk-SS vertical electron transport, as has been discussed previously elsewhere.[11,14]

The estimated rates in the 2D TI $Bi_2Se_3$ are ~1.7 and ~2.8 eV ps$^{-1}$, whereas in the 3D TI $Bi_2Se_3$, they are ~1.7 and ~3.5 eV ps$^{-1}$. We note here that because the electron relaxation dynamics occurs through both the bulk states and Dirac SS simultaneously, the two-stage relaxation dynamics is less distinguishable in thinner films due to stronger coupling between the opposite surfaces of the film [Figure 4(a)]. The lower rate of the relaxation dynamics through the bulk states in the 2D TI $Bi_2Se_3$ compared to that in the 3D TI $Bi_2Se_3$ points hence to the dominating relaxation trend occurring through the Dirac SS with decreasing film thickness. Consequently, the rates estimated for the 3D TI $Bi_2Se_3$ seem to be more reliable and being closer to actual ones [Figure 4(c)]. For the same reason, the effect of the Dirac SS3 on the relaxation dynamics is expected to be masked by the bulk states for low pumping powers [Figure 4(a)], whereas the effect of the Dirac SS4 and SS3 is appeared predominantly with increasing photoexcited carrier density [Figure 4(c)].

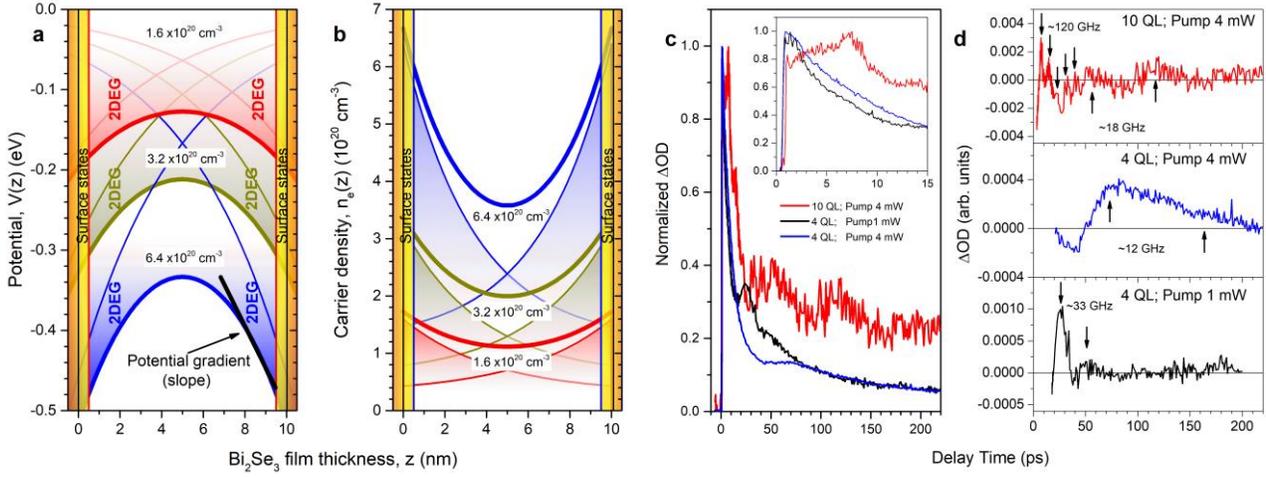

**Figure 5.** Surface potential, spatial carrier distribution, and pump-probe traces of FCA in the 3D TI $Bi_2Se_3$ film. (a) and (b) The surface potential profile and carrier density distribution simulated for the 10 QL thick 3D TI $Bi_2Se_3$ film, respectively. Different color curves correspond to different carrier densities, as indicated. The thin curves present the effect of the individual surfaces, whereas the broad curves present the joint effect of two surfaces. The 2DEG and the potential gradient (slope) are shown in (a). (c) and (d) The pump-probe traces of FCA measured for the 3D and 2D TI $Bi_2Se_3$ under conditions, as indicated, and the corresponding extracted oscillatory parts. Inset in (c) shows the short delay time part of the same traces. The peaks of the oscillatory behavior associated with acoustic Dirac plasmons and the corresponding frequencies are indicated in (d).

The about two times lower cooling rate of electrons in the Dirac SS as compared to that in the bulk states points to the different types of electron-phonon coupling: the inelastic-type in the bulk states and the quasi-elastic-type in the Dirac SS [Figure 4(b) and (d)].[11] Nevertheless, the electron relaxation rates in the 3D TI $Bi_2Se_3$ weakly depends on whether the pumping power is below or above the critical pumping power for the dynamic topological phase transition. This behavior is not surprising since the Fröhlich interaction in the ideal case is independent of photoexcited carrier density.[17]

Because the substantial downward band bending near the surface in unintentionally $n$-doped $Bi_2Se_3$, the SQW is created, confining electrons to the surface and thus forming 2DEG [Fig. 5(b)].[12,13,14] The bending is known to cause a surface potential, the profile of which along the surface normal ($z$-axis) causes the depletion layer width ($z_d$) and can be estimated for low carrier densities by solving the corresponding Poisson equation to yield[18,19,21]

$$V(z) = -\frac{en_e}{2\varepsilon\varepsilon_0}z^2, \qquad (1)$$

where $e$ is the electron charge; $\varepsilon_s$ and $\varepsilon_0$ are the static dielectric constant and the permittivity of free space, respectively. For unintentionally $n$-doped $Bi_2Se_3$, the actual $z_d$ may be a few tens of nanometers.[13,15,18,19] Upon optical pumping, the carrier density significantly increases, thus deepening the SQW, shortening $z_d$, enhancing the electron confinement in a close vicinity of the surface, and balancing the corresponding Fermi energy of the 2DEG at quasi-equilibrium: $E_F = (\hbar^2/2m_e^*) \times (3\pi^2 n_e)^{2/3}$, where $m_e^*$ is the electron effective mass and $\hbar$ is the reduced Planck constant.[20,21] The resulting spatial distribution of carriers in the 2DEG below $E_F$ can be obtained using the Thomas-Fermi screening model:[20,21]

$$n_e(z) = \frac{2\varepsilon_s\varepsilon_0 E_F}{3e^2 z^2} \approx [n_e(0)]^{2/3} \times \frac{3\varepsilon_s\varepsilon_0 \hbar^2}{e^2 m_e^*}\left(\frac{1}{z}\right)^2, \qquad (2)$$

where $n_e(0)$ is the carrier density of the 2DEG right in front of the Dirac SS.

Figure 5 shows the numerical modeling of the depth dependence of the surface potential and carrier density for the three carrier densities photoexcited in the 3D TI $Bi_2Se_3$ film using the corresponding material parameters[20] and taking into consideration the effect of an individual surface and two identical surfaces of the film. One can see that both the surface potential extension and carrier density distribution at the individual surface of the film is of the order of the film thickness (~10 nm). Furthermore, both the dependencies become shorten with increasing photoexcited carrier density, thus increasing the depth of the SQW and carrier density in it. If the joint effect of the two identical surfaces of the film is treated, the carrier density at surfaces increases owing to the deeper surface electrostatic potential, nevertheless it becomes high enough in the middle of the film as well. The latter behavior makes the film poorly metallic and causes the electrostatic coupling between opposite surfaces. The corresponding surface potential gradient [$\nabla V(z)$], equally the depletion electric field strength, $E_{dep} = -\nabla V(z)$,[21] also increases with increasing the photoexcited carrier density.

As we mentioned above, non-magnetic perturbations cannot open a true band gap at the Dirac SS nodes.[2] However, owing to the potential gradient that breaks the inversion symmetry along the $z$-axis, the 2DEG experiences a Rashba spin-splitting, where electrons with opposite spins are separated in energy,[3] thus introducing a net magnetic moment in a close vicinity of the surface.[12,14] Specifically, the effective Hamiltonian for the 2DEG subjected to strong SOC is known to be expressed as,[15,16]

$$H = \frac{\hbar^2 \mathbf{k}^2}{2m_e^*} - \mathcal{H}_R(\mathbf{k}), \qquad (3)$$

where the first term is the kinetic energy term and $\mathcal{H}_R(\mathbf{k}) = \alpha_R(k_x\sigma_y - k_y\sigma_x)$ is the Rashba term with $\mathbf{k}$, $\alpha_R$ and $\sigma_i (i = x, y, z)$ being the wavevector, the Rashba parameter, and the Pauli matrices, respectively. The parallel and perpendicular directions with respect to the plane of the $Bi_2Se_3$ film are hence $\mathbf{k}_\parallel = (\mathbf{k}_x, \mathbf{k}_y)$ and $\mathbf{k}_\perp = \mathbf{k}_z$, respectively. Once the $E_{dep}$ field is applied along the z-axis, the lack of inversion symmetry leads to the Rashba spin-splitting appearing in the $\mathbf{k}_\parallel$ space. The corresponding in-plane Rashba magnetic field in energy units is defined as,[22]

$$\mathbf{B}_R = \frac{2\alpha_R}{g\mu_B}(k_x\sigma_y - k_y\sigma_x). \quad (4)$$

where g and $\mu_B$ are the g-factor and the Bohr magneton, respectively. In general, it is known that in-plane magnetic order causes shifting of the spin-splitting bands in momentum space, whereas out-of-plane magnetic order causing gap opening at the Dirac SS node is usually expected to be negligible.[5] This ideal picture presents well the Rashba effect dynamics occurring at low and moderate photoexcited carrier densities. However, because the Rashba effect linearly depends on the electric field strength and hence on the photoexcited carrier density,[16,23] the effective magnetic field associated with the Rashba spin-splitting in the $Bi_2Se_3$ bulk is expected to reveal significant magnetic anisotropy with increasing the photoexcited carrier density. The resulting out-of-plane spin polarization is assumed hence to be caused by anisotropy of surface potential scattering and non-parabolicity of the energy spectrum,[24] both being induced by strong inter-surface electrostatic coupling in the film.

The latter behavior is also well consistent with frequency variations of oscillation modes observed in the FCA traces and associated with coherent acoustic Dirac plasmons [Fig. 5(c) and (d)].[14] In the 2D TI $Bi_2Se_3$ (gapped SS), the frequency of ~33 GHz observed for low pumping powers corresponds to the transverse electric mode (in-plain magnetic oscillations), whereas this frequency is decreased to ~12 GHz with increasing pumping power, corresponding to the transverse magnetic mode (out-o-plane magnetic oscillations).[11,14] As a gap at the Dirac SS node is opened in the 3D TI $Bi_2Se_3$, the acoustic Dirac plasmon appears through both modes with frequencies ~120 and ~18 GHz, respectively [Fig. 5(d)]. Thus, the induced out-of-plane magnetization breaks the time-reversal symmetry in Dirac SS, making the surface Dirac fermions massive and resulting in opening a gap at the Dirac SS node.[5] This behavior causes the topological phase transition persisting as long as the photoexcited electron population in the film remains sufficiently high.

In summary, the experimental findings presented in this letter demonstrate light-controlled dynamic opening of a gap at the Dirac SS node in thin films of the 3D TI $Bi_2Se_3$ induced indirectly by the dynamic Rashba effect occurring in the film bulk with increasing the photoexcited carrier density (optical pumping power). The observed effect is expected to be used to control the operating current modulation regimes in optoelectronic nanodevices based on topological insulators.

**Supporting Information:** 1. Preparation of the samples and their characterization. 2. Experimental setup description. 3. Full set of the corresponding pump-probe traces measured at certain probing wavelengths and pumping powers.

## AUTHOR INFORMATION
### Corresponding Authors
**Yuri D. Glinka** − *Guangdong University Key Lab for Advanced Quantum Dot Displays and Lighting, Shenzhen Key Laboratory for Advanced*

*Quantum Dot Displays and Lighting, Department of Electrical and Electronic Engineering, Southern University of Science and Technology, Shenzhen 518055, China; Institute of Physics, National Academy of Sciences of Ukraine, Kiev 03028, Ukraine;* orcid.org/0000-0002-2267-0473; Email: yuridglinka@yahoo.com

**Tingchao He** − *College of Physics and Energy, Shenzhen University, Shenzhen 518060, China;* orcid.org/0000-0003-1040-0596; Email: tche@szu.edu.cn

**Xiao Wei Sun** − *Guangdong University Key Lab for Advanced Quantum Dot Displays and Lighting, Shenzhen Key Laboratory for Advanced Quantum Dot Displays and Lighting, Department of Electrical and Electronic Engineering, Southern University of Science and Technology, Shenzhen 518055, China; Shenzhen Planck Innovation Technologies Pte Ltd., Shenzhen 518112, China;* orcid.org/0000-0002-2840-1880; Email: sunxw@sustech.edu.cn

**Acknowledgements**
This work was supported by the National Key Research and Development Program of China administrated by the Ministry of Science and Technology of China (Grant No. 2016YFB0401702), the Guangdong University Key Laboratory for Advanced Quantum Dot Displays and Lighting (Grant No. 2017KSYS007), the National Natural Science Foundation of China (Grant Nos. 11574130 and 61674074), the Development and Reform Commission of Shenzhen Project (Grant No. [2017]1395), the Shenzhen Peacock Team Project (Grant No. KQTD2016030111203005), and the Shenzhen Key Laboratory for Advanced Quantum Dot Displays and Lighting (Grant No. ZDSYS201707281632549). The authors acknowledge J. Li for help with the laser system operation and S. Babakiray for growing the $Bi_2Se_3$ samples by MBE (under supervision of D. Lederman) using the West Virginia University Shared Research Facilities.

**Author contributions**
Y.D.G. modified and tested the transient absorption spectrometer, built the experimental setup, performed optical measurements, and treated the optical experimental data. The optical measurements were performed in the laboratory hosted by T.H. All authors contributed to discussions. Y.D.G. analyzed the data and wrote this paper. X.W.S. guided the research and supervised the project.